Ullah, F., Johannes Raft, A., Shahin, M., Zahedi, M., and Babar, M.A (2017) 'Security Support in Continuous Deployment Pipeline', *Proceedings of 12th International Conference on Evaluation of Novel Approaches to Software Engineering*

# Security Support in Continuous Deployment Pipeline

Faheem Ullah[1], Adam Johannes Raft[2], Mojtaba Shahin[1], Mansooreh Zahedi[2] and Muhammad Ali Babar[1,2]

[1]CREST – Centre for Research on Engineering Software Technologies, The University of Adelaide, Adelaide, Australia
[2]CREST – Centre for Research on Engineering Software Technologies, IT University of Copenhagen, Copenhagen, Denmark
{faheem.ullah, mojtaba.shahin, ali.babar}@adelaide.edu.au, {adamraft, mzah}@itu.dk

Abstract: Continuous Deployment (CD) has emerged as a new practice in the software industry to continuously and automatically deploy software changes into production. Continuous Deployment Pipeline (CDP) supports CD practice by transferring the changes from the repository to production. Since most of the CDP components run in an environment that has several interfaces to the Internet, these components are vulnerable to various kinds of malicious attacks. This paper reports our work aimed at designing secure CDP by utilizing security tactics. We have demonstrated the effectiveness of five security tactics in designing a secure pipeline by conducting an experiment on two CDPs– one incorporates security tactics while the other does not. Both CDPs have been analysed qualitatively and quantitatively. We used assurance cases with goal-structured notations for qualitative analysis. For quantitative analysis, we used penetration tools. Our findings indicate that the applied tactics improve the security of the major components (i.e., repository, continuous integration server, main server) of a CDP by controlling access to the components and establishing secure connections.

## 1 INTRODUCTION

Continuous Deployment (CD) is a software development practice which enables an organization to deploy software to customers continuously, automatically and reliably (Claps et al., 2015, ElectricCloud, 2016). A number of innovative organizations such as Facebook, Microsoft, and IBM adopted CD to deliver values to their customers frequently. CD brings several benefits to an organization (Anderson, 2014). These benefits include reducing developer's effort, improving the quality of software, and reduced cost (Anderson et al., 2014, Chen, 2015). Continuous Deployment Pipeline (CDP) is the core concept to successfully implement CD practice (contributors, 2016, Humble and Farley, 2010, Phillips A, 2015). CDP automatically transfers code changes from a repository to a production environment. Furthermore, CDP enables the team members to always keep an eye on every aspect (e.g., build, deploy, test etc.) of the system, and get a quick feedback on deployed software. A CDP also promotes collaboration between various groups of developers working together to fix bugs and issues and deliver the software by improving the visibility of changes (Fowler, 2013). A CDP is a collection of stages (e.g., build, package, and test) supported by tools (GitHub, Jenkins, AWS etc.) and technologies for enabling continuous and automated deployment of changes into production. The number and nature of stages involved in CDP vary from organization to organization (Adams and McIntosh, 2016). Similarly, the tools and technologies incorporated for implementation of CDP also vary from project to project and organization to organization.

Security of software supply chain is becoming important because of the involvement of several direct and indirect participants in the process (Ellison et al., 2010). In order to ensure a secure supply of software, each phase (initiation, development, deployment, maintenance and disposal) of software supply chain needs to be protected from malicious attacks. Being the last portion of the supply chain, deployment pipeline needs to be fully secure (Bass et al., 2015). However, the reality is contrary to this. Different users from various teams (e.g., development, operation, and testing) have the same level of access to various resources on the pipeline which gives unnecessary access and paws way for malicious activities (Rimba et al., 2015). Continuous Integration (CI) server, an important part of a CDP, generally has a monolithic design which enables an attacker (who breached a single part of the code) to have access to all parts of the code and so gain an overall control of the entire process (Bass et al., 2015). Securing a CDP is a challenging task due to the variety of tools involved with each having its own security requirements (Bass et al., 2015).

It is asserted that if the components of a CDP and the communication among them are secure, then the whole CDP will be secure (Bass et al., 2015, Rimba et al., 2015). Hence, we propose the use of five security tactics for protecting CDP from malicious attacks by addressing the security requirements of the three major components (i.e., repository, main server, and CI server) of the CDP. The primary focus of our security tactics is to ensure controlled access to these components. We demonstrate the effectiveness of our security tactics by comparing two CDPs – one that incorporates our proposed tactics and other that does not. Our results show that security tactics ultimately lead to enhancing the security of the entire CDP. It is worth mentioning that both academia and industry refer to CDP and CI server also as continuous delivery pipeline and automated build server respectively. Therefore, these terms are used interchangeably in the rest of the paper.

The rest of the paper is organized as follows: Section 2 discusses CDP, its security in the light of existing literature, and motivation for this work. Section 3 includes an overview of our implemented CDPs, security risks identified for each of the three components, and presents our approach for eliminating identified risks through the incorporation of our proposed security tactics. Section 4 presents analysis and results from the qualitative and quantitative evaluation of the effectiveness of security tactics. Section 5 provides a discussion on the results and limitations of our approach. Section 6 concludes the work and identifies some future research directions.

## 2 RELATED WORK

Sufficient research exists on the identification and categorization of software security risks. Reviewing such literature gives us an idea of possible permutations inside a software system. (Landwehr et al., 1993) classify security flaws based on how, when and where they are introduced into the system. Based on this logic, security flaws are categorized into three categories: Genesis (intentionally, unintentionally etc.), Time of Introduction (during development, maintenance, or operation etc.) and Location (hardware or software). ((Langweg, 2004) categorize attacks that software applications can come across. According to this classification, attacks are divided into three categories: Location (input), Cause (processing), and Effect (output). (Aslam et al., 1996) present the classification of security faults in Unix Operating System to highlight various types of security faults. Similarly, several organizations also highlight security risks in software. Open Web Application Security Project (OWASP)[1] created a list of top 10 vulnerabilities (e.g. injection, broken authentication & session management, and missing function-level access control etc.) for web applications. In 2011, Common Weaknesses Enumeration (CWE)[2] also published a list of 25 software errors (missing authentication, missing authorization, incorrect authorization etc.) that can lead to serious losses.

(Bass et al., 2015) explore various scenarios of subverting a pipeline that includes deployment of an invalid image, deployment of an image without being passed through a complete pipeline, and unauthorized environment (e.g. development) having direct access to the production environment. Authors propose steps for securing the pipeline that includes: (1) identification of security requirements of the pipeline; (2) differentiating between trustworthy and untrustworthy components of the pipeline; (3) decomposition of untrustworthy components of the pipeline; (4) modification of untrustworthy components to let the trustworthy components perform critical operations. The proposed process for securing the deployment pipeline is aimed at making trustworthy components of the pipeline mediate access to the actual building and deploying activities. Accessing sensitive data or functions only through trustworthy components improves the security of the pipeline by preventing untrustworthy components from accessing sensitive functions. The devised process does not fully secure the pipeline but hardens it to a certain level.

(Rimba et al., 2015) highlight several security requirements of CDP that include: (1) different roles (e.g. development team, operation team etc.) should have different levels of access (2) in order to prevent malicious code end up being deployed in production, CDP should not be miss-configured or compromised in any way and (3) testing and production environments should be fully isolated. Authors demonstrate the suitability of their proposed approach (Design Fragments) by securing a CDP to satisfy its security requirements. In order to address first security requirement, authors utilize existing security mechanisms of Amazon Web Service (AWS) and CI server (Jenkins) to assign different access levels to different users. For second security requirement, AWS buckets (*codeBucket, credsBucket, imageBucket, and configBucket*) have been protected by allowing only Jenkins to have access to them. Authentication enforcer design fragment has been inserted between Jenkins and buckets, and devised tactics are leveraged to make required connections or disconnections for separating Jenkins from trusted components. Using execution domain pattern, authors define three logical execution domains (testing, production, and shared) for isolation of testing and production environments. Assurance Case Analysis has been performed to verify that devised tactics fully address second and third security requirement of the CDP.

---

[1] https://www.owasp.org/index.php/Category:OWASP_Top_Ten_Project#tab=OWASP_Top_10_for_2013

[2] http://cwe.mitre.org/top25/

(Gruhn et al., 2013) analyse CI from the security perspective to identify possible security threats. This study relates to our work as it also identifies a class of threats related to build server. Build Server executes a build job in four steps: (1) Version Control System (VCS) checkout (2) Build preparations (3) Builder runs (4) Notification. Each step is vulnerable to various kinds of malicious attacks such as exploiting symbolic links (Ko et al., 1994), Denial of Service attack, Thompson's trusting trust attack (Thompson, 1984). These threats are eliminated by encapsulating build job through virtualization. The CI system restores build server to its original clean form after every build process and thereby, protects build server from malicious attacks.

This related work section gives us an insight into CDP security risks through investigation of security taxonomies, findings of various security organizations and related research works. From these findings, it can be extracted that CDP is subjected to a vast majority of security threats. In existing literature, some studies (Bass et al., 2015, Rimba et al., 2015) focus on access control while some (Gruhn et al., 2013) focus on virtualization for securing build server. Our approach leverages both access control measures and virtualization for securing the pipeline. Similarly, existing approaches are primarily focused on securing build server (which is one component of the CDP) while our proposed tactics secure three main components (repository, main server, and build server) of the CDP. Most importantly, existing approaches are evaluated using only qualitative analysis. We evaluate the effectiveness of our proposed security tactics using both qualitative and quantitative analysis.

# 3 APPROACH

First, this section briefly describes our CDP and shows how basic components of our implemented CDPs collaborate with each other. Then, the CDP is analysed from the security perspective to identify the basic security risks in the CDP. The identification of these security risks helps us in designing our security tactics. Finally, we describe proposed security tactics for improving the security of our CDP.

## 3.1 Overview of CDP

The three main components of our CDP and the relation between them is shown in Fig – 1. The repository is the place where developers commit their developed code. CI server is responsible for testing and building the code committed to the repository. In case commit of a developer breaks the commit of another developer, then corresponding developer is informed. If the build is successful then the code is deployed in the main server.

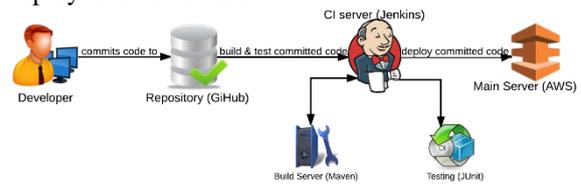

Figure 1: Continuous Deployment Pipeline (CDP).

The components of the CDP, tools used for implementation of the corresponding components, and their versions are shown in Table – 1. For the purpose of comparison, two CDPs are implemented – one incorporates the security tactics (Secure CDP) and other does not (Non-secure CDP). In both CDPs, except GitHub, all other components run on an AWS instance with Ubuntu as OS.

Table 1: Components of CDP.

| Component | Tool | Version |
|---|---|---|
| Repository | GitHub | 1.9.1 |
| CI Server | Jenkins | 1.656 |
| Test | JUnit | 4.11 |
| Build Server | Maven | 2.2.1 |
| Web Server | Tomcat | 7.0.52.0 |

## 3.2 Security Risks in CDP

One of the major challenges in implementing CDP is dealing with security risks (Bass et al., 2015, Rimba et al., 2015). Before devising any approach for securing CDPs, it is imperative to first identify and understand these security risks faced by various

components of the CDP as summarized in Table – 2 and described in the followings:

### 3.2.1 Security Risks in Repository

Repository (GitHub) of our CDP is a standalone component that does not borrow or lend security to any other component. Since a password is the protection criterion that repository uses to authenticate developers, therefore, password implementation needs to be of high strength (Gaw and Felten, 2006). Secondly, a user with an access to the GitHub account has total control over all other repositories associated with that account. This total control includes deleting individual repositories and accepting a push or pull request for others. If such a request for a malicious user is accepted, then this user may initiate malicious activities and may accept requests for other malicious users.

### 3.2.2 Security Risks in Main Server

Access to the Main server (AWS) should be authenticated and authorized. Although a high strength password solution is a fairly secure option, but sometimes average password solutions are implemented which gives an opportunity to social engineers to breach password and get unauthenticated access to resources (Tari et al., 2006). In addition to password protection, an additional security measure needs to be taken to enhance the authentication process for the Main server. Similarly, once authenticated, a user gets full access to the instance including the OS. A mechanism is required to restrict the access to resources on the Main server.

### 3.2.3 Security Risks in CI server

CI server (Jenkins) also faces serious security threats. A security failure can cause malicious injection in a VM instance (with Jenkins inside it) while it is running. It is important to ensure that before starting a new build process, CI server should be in a clean state (Gruhn et al., 2013). Secondly, the default installation of Jenkins gives free access to everyone. A mechanism is needed to assign a role to each user which specifies the access rights of the user (Sandhu et al., 1996). Such a mechanism would enable the administrator to control who can create, modify and delete pipelines.

Table 2: Security Risks in Key Components of CDP.

| Component | Security Risks |
|---|---|
| Repository (GitHub) | Uncontrolled access |
| Main Server (AWS) | Poor authentication mechanism |
|  | Uncontrolled access |
| CI server (Jenkins) | Starting build process with previously infected state |
|  | Uncontrolled access |

## 3.3 Proposed Security Tactics

After a thorough analysis of the security threats posed to various components of the CDP, five Security Tactics (ST) are devised to eliminate identified threats and secure the pipeline against malicious activities. These security tactics are:

*1. Securing repository through controlled access to get hold over who can commit to certain branches of the repository*
*2. Securing connection to the main server through use of private key over Secure SHell (SSH)*
*3. Using roles on the main server to control access via leveraging AWS Identity and Access Management (IAM) ecosystem[3]*
*4. Setting up the CI server to start up a Virtual Machine (VM) with a clean state by leveraging Jenkins VM plug-in (Jenkins, 2013)*
*5. Using Jenkins roles plug-in (Jenkins, 2016) for assigning roles on the CI server to control who can create, modify and delete pipelines*

First two tactics are incorporated in both the CDPs (Secure CDP and non-secure CDP) while rest of the three tactics are only incorporated in the secure CDP as shown in Fig – 2. Each of the tactics is further explained in the following sub-sections.

---
[3] https://aws.amazon.com/documentation/iam/

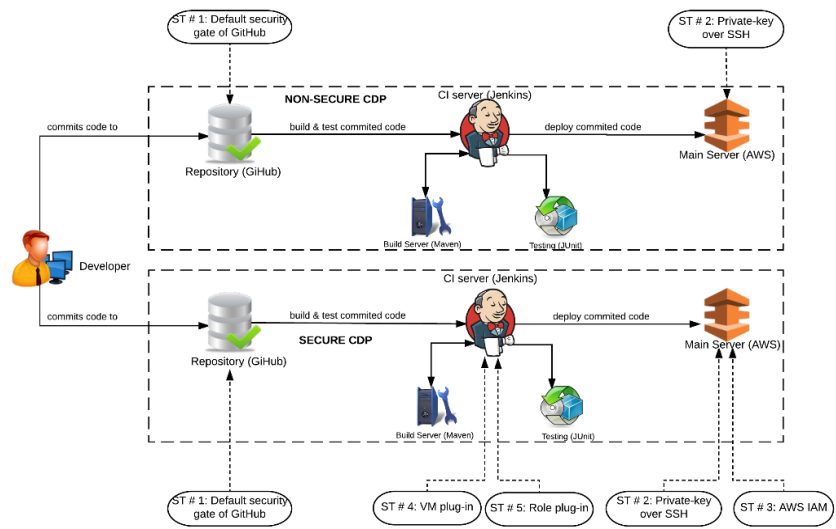

Figure 2: Secure & non-secure CDP with incorporated security tactics.

### 3.3.1 Controlled Access to Repository

The Repository is the starting point of the CDP and if its security is breached, then the security of the entire CDP becomes vulnerable. GitHub allows developers to commit code to the project by adding them to "Collaborators". In order to have control over who can commit code or create and delete individual repositories, default security gate of GitHub is utilized. This enables the administrator or particular user with assigned rights to accept or reject a commit request. Each time a user makes a push request to commit code, the administrator of the repository has the authority to accept or reject the request. Applying this approach before accepting any commit request enables the administrator to ensure that user or his activity is not malicious. Sometimes, it may not be possible to have the administrator to make an actual pull for every commit due to a high number of commit requests. However, there exist several solutions to address this issue. For example, if the server is propriety Git server then *Gitolite[4]* is a possible solution.

### 3.3.2 Enhanced Authentication Mechanism for Main Server

In addition to username and password, private key over SSH (Ellingwood, 2014) is leveraged by the Main server to keep AWS instance safe from an insecure connection. Username and password give access to AWS interface where instances can be manipulated but username and password cannot enable a user to connect to an instance. In order to connect to an AWS instance, a private key over SSH is required. This additional protection through private key over SSH enhances authentication process and ensures that no malicious user is connected to an AWS instance.

### 3.3.3 Controlled Access to Main Server

Having only authentication mechanism means all users will have the same kind of access rights, which is problematic. In order to allocate particular access rights to particular users, the concept of roles is introduced. AWS Identity and Access Management (IAM) ecosystem can be utilized to enable an administrator to control access of users to AWS instances and ecosystem and allocate access rights based on the particular role of the user. For example, the administrator can control which user can change the settings of a firewall.

### 3.3.4 Clean CI Server VM Image

Utilizing VM plug-in in Jenkins protects VM from outside malicious access (Gruhn et al., 2013). Every time a Jenkins is asked to build, it fires up a VM with a Jenkins inside it. Since the Jenkins is inside the VM that performs the build, therefore, Jenkins instance is not vulnerable to malicious activity. When the build process gets finished, VM is shut down and the Jenkins instance inside this VM is destroyed. Next time, when a Jenkins is asked to build, a new VM with a new Jenkins instance is created to start the new clean build. Fig – 3 highlights the significance of VM plug-in by showing the difference between states of a CI server in the presence and absence of VM plug-in.

---
[4] https://git-scm.com/book/en/v1/Git-on-the-Server-Gitolite

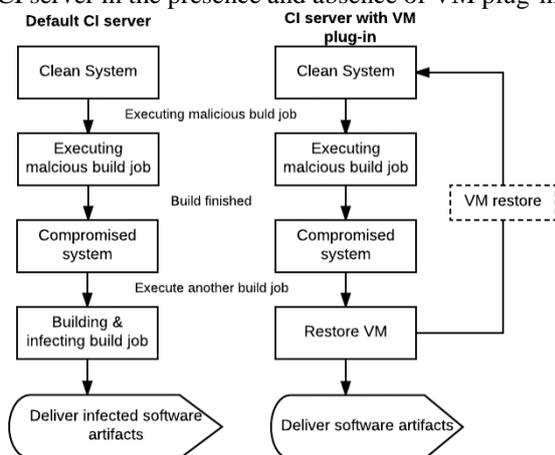

Figure 3: States of CI server with and without VM plug-in.

### 3.3.5 Controlled Access to CI Server

With Jenkins' roles plug-in, it is possible to create global roles, project roles, slaves' roles and user roles ('Role Strategy Plugin. Available at https://wiki.jenkins-ci.org/display/JENKINS/Role+Strategy+Plugin [Last Accessed: 24th Oct, 2016],'). Here, we are particularly interested to leverage this plug-in for enabling the administrator to have a control over the activity of a user. Using roles plug-in, administrator assigns roles to each user based on his particular role. Such an assignment of role would decide access rights of the user. For example, an administrator may restrict one user from creating, modifying or deleting a pipeline but may allow another user to perform these tasks.

## 4 ANALYSIS AND RESULTS

This section analyses the implemented CDPs both qualitatively and quantitatively to investigate whether the proposed tactics enhance the security of secure CDP.

### 4.1 Qualitative Analysis of CDPs

We use Assurance Case with Goal Structuring Notation (GSN) for qualitative assessment of the effectiveness of proposed security tactics. Assurance Case is a qualitative testing technique where evidence is organized into an argument to show to a certain interested party that a certain claim regarding the system holds true (John Goodenough, 2007). In Assurance Case technique, a claim about a system is established and supported by objective evidence. Sometimes safety arguments within safety cases communicated via free text are unclear and create misunderstanding among various stack holders. It is always efficient and easily understandable to present assurance case in graphical form rather than textual form. For this purpose, GSN (Kelly and Weaver, 2004) is used to properly communicate arguments in an assurance case through graphical notations. In GSN, elements are linked together to form a goal structure and while supporting arguments, goal structure is successively broken down into sub-goals until these small goals can be directly supported via evidence (Kelly and Weaver, 2004).

We aim to secure three basic components (Repository, Main Server, and CI Server) of a CDP. We will analyse whether our proposed security tactics meet the security requirements of these three components of a CDP. If we demonstrate that the proposed tactics properly meet the security requirements, then it can be shown our security tactics improve a CDP's security.

From the security perspective, the repository requires controlled access, which means not all users, should have full rights to access every resource or perform any operation at the repository. Security requirements of the Main server can be broken down into two parts: firstly, every user should be properly authenticated before allowing him access to the Main server; and secondly access to resources or authority to perform operations should be authorized. The security requirements of the CI Server can also be broken down into two parts: firstly CI server should be in the clean state before starting a new build process; and secondly access to CI server should be controlled so that the principle of least privilege (Sandhu and Samarati, 1994) can be realized. We make an assurance case as shown in Fig – 4 to argue that our proposed tactics satisfy the security requirements of the CDP. We claim that our CDP is secure because three of the major components (repository, main server, CI server) of the CDP are secure. The repository is secure because access to the repository is totally controlled. First, a user is authenticated through his credentials (username and password). After being authenticated, default security gate of GitHub is leveraged which enables the administrator to decide about user's privileges. The mechanism allows the administrator to keep a check on who is committing code and prevents a common user from allowing an attacker to commit his malicious code. This security measure also provides an additional protection to Java files, JUnit files, and Maven files because they do not have their own security mechanism rather rely on repository's security mechanism. Next, we claim that Main Server is secure. Main Server will be secure if users accessing the Main Server are authenticated and each user has specific rights according to his role to access or perform particular operations. In order to access an AWS instance on Main Server, in addition to username and password, a user needs private-key over SSH. This additional security measure ensures that even if a malicious user breaches the normal password security system, he can't connect to the AWS instance as he would require a private key for connecting and manipulating an AWS instance. Similarly, the second requirement of the Main server is addressed by utilizing the AWS IAM service that enables an administrator to assign specific access rights to users according to their roles. Finally, we claim that CI server is also secure. This claim is supported by two arguments. In order to ensure that CI server is in a clean state before starting a build process, we are leveraging the VM plug-in, which protects Jenkins instance from malicious attacks and ensures that CI server remains in the clean and non-infected state. Since the security of CI server requires controlled access to CI server, role plug-in is leveraged to enable an administrator to assign roles to various users according to their particular roles.

Since our qualitative analysis demonstrates that the proposed security tactics satisfy the security requirements of the CDP, therefore, we can establish that our proposed security tactics contribute to improving the security of the CDP.

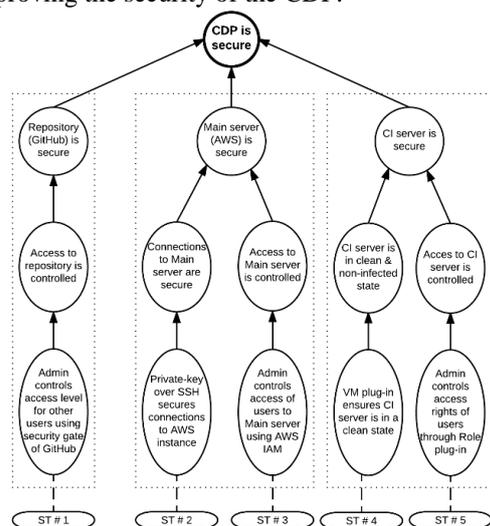

Figure 4: CDP Assurance Case.

---

[5] https://www.qualys.com/forms/freescan/owasp/

## 4.2 Quantitative Analysis of CDPs

For quantitative evaluation of the effectiveness of security tactics, two scanning tests are performed. These scanning tests launch various kinds of attacks on the application to find vulnerabilities and assess the security level of the application.

The first of these tests is the Qualys OWASP Scan[5] that is normally practiced to see whether a web application works according to the security standards set by OWASP against online attackers. Qualys OWASP scan helps understand and identify vulnerabilities and support in fixing these vulnerabilities. Scanning engine is intelligently designed to perform specific scanning tasks and avoid unnecessary vulnerability checks. Qualys scanning methodology follows the same steps as an attacker would follow (Qualys, 2015). The basic steps of the scanning process include: (1) checking if the host to be scanned is alive and running; (2) checking if host is using some firewalling; (3) identifying all open TCP and UPD ports; (4) checking which operating system is used by host; (5) identification of services running on open TCP or UDP ports; (6) starting actual non-intrusive vulnerability assessment (Qualys, 2015).

The second scanning tool is OWASP Zed Attack Proxy (ZAP)[6] scanner that is a free security scanner for finding vulnerabilities in web applications. ZAP has two kinds of scanners: Active and Passive (ZAP, 2015). Active scanner performs a wide range of known attacks on the host to find vulnerabilities. The active scanner cannot detect logical vulnerabilities such as broken access control. In addition to active scanning, it is always beneficial to perform manual penetration testing too. Passive scanner constantly examines requests and responses to detect a certain type of vulnerabilities. ZAP also has fuzzing capability to identify vulnerabilities that are more settled, which active and passive scanners cannot identify. In this work, we only focus on automatic attacks to assess the security aspects of CDPs.

Primarily, these tools focus on web aspect of penetration testing. The two important components of CDPs (GitHub and Jenkins) have a public interface in the form of a website. Tomcat, which hosts Jenkins, has a public interface and so does the dashboard that controls AWS instances. Keeping in view that CDP has public web interfaces, these tools are best available tools for quantitative assessment of the security level of CDPs.

---

[6] https://www.owasp.org/index.php/OWASP_Zed_Attack_Proxy_Project

### 4.2.1 Repository (GitHub)

We mentioned that a single repository is used with both the CDPs, hence, the security level of the repository cannot be compared. Instead, these tests enable us to find the vulnerabilities and their severity.

OWASP scan found 105 vulnerabilities in the repository as shown in Fig – 5a. The majority of the vulnerabilities are related to Denial-Of-Service (DoS) attacks, Internet Control Message Protocol (ICMP) timestamp, path, and password-completion. DoS attacks do not pose any direct threat to the security of GitHub as these issues can affect communication to and from pipeline but cannot directly infect the pipeline. As a matter of the best practice, ICMP timestamp issues can be addressed via several available techniques (Singh et al., 2003, Security, 2016), but these issues do not have any significance in relevance to a CDP's security. There are several path-based vulnerabilities as well which again does not pose any serious threat to a CDP's security. These path-based vulnerabilities give the attacker some information about folder structure on the server, which can be used for guessing the structure of other folders on a server. Most browsers have auto-password completion feature, which is a serious issue. It means that retrieving such a password from the browser would enable an attacker to access CDP and inject malicious software, which will be a total breach of security.

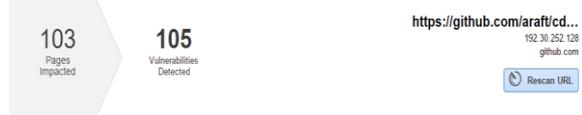

Figure 5a: OWASP Scan Result for Repository (GitHub).

ZAP scan found several vulnerabilities categorized into eight groups as shown in Fig – 5b. Identified vulnerabilities are related to settings of cookie, usage of JavaScript, content caching, IP disclosure and password auto-completion. Setting cookie without the *secure* flag and *HTTPOnly* flag makes it possible to access cookie via non-encrypted connection and using JavaScript respectively. It does not have much to do with a CDP's security and can be easily fixed too. The results show that about 6618 vulnerabilities of using JavaScript for another domain. Not all the cases have been checked but the ones that are checked come from GitHub subdomain *asssets-cdn.github.com* which makes it a non-issue in relevance to a CDP's security. There are around 3683 cases (vulnerabilities) where HTTP allows browser or proxy to cache contents, which again is not relevant to the security of CDP. There are also cases of displaying private IP in HTML response code that can be mitigated via Load Master Content Rule (KEMP, 2016) or similar strategies depending upon the type of server. This vulnerability is also not directly related to the security of CDP. Like OWASP scan, password auto-completion vulnerability is detected by ZAP scan too, which poses a serious threat to the security of CDP.

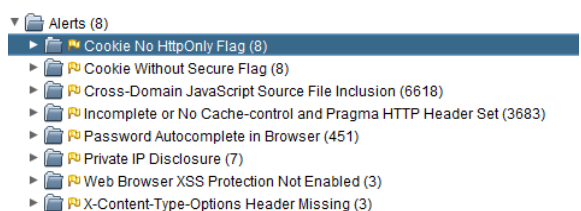

Figure 5b: ZAP Scan Result for Repository (GitHub).

### 4.2.2 Main Server (AWS)

Scanning tests are separately applied on Main servers for secure and non-secure CDPs.

**Main Server of Secure CDP**

As shown in Fig – 6a OWASP scan found three vulnerabilities in the Main server of secure CDP. Vulnerabilities found by this scan are related to cookies, which identifies that *secure* flag and *HTTPOnly* flag are not set. If these flags are not set, it may allow the browser to communicate via a non-encrypted channel and a client side script would be able to read a cookie. Hence, such vulnerabilities do not affect the security of CDP. Additionally, these issues can be easily fixed.

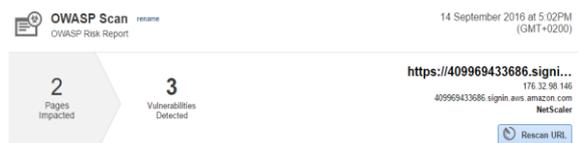

Figure 6a: OWASP Scan Result for Main Server (AWS) of Secure CDP.

ZAP scan found around 26 vulnerabilities of six different types. Results obtained from ZAP scan are shown in Fig – 6b. Similar to OWASP scan, the majority of vulnerabilities are relevant to cookies. Apart from that, issues relevant to content caching and cross-site scripting are also identified. The list of vulnerabilities shows that X-Frame-Options Header is not added. This allows an attacker to inject multiple transparent layers in HTTP page for deceiving a user. Most modern browsers have this feature and this issue can be easily fixed. As mentioned previously, the issue of content caching is hardly relevant to the security of CDP. Similarly, the Anti-MIME-Sniffing header X-Content-Type-Options can be easily set to 'nonsniff'. Further results indicate that XSS protection is not enabled which can be enabled by setting the X-XSS-protection HTTP response header to '1'.

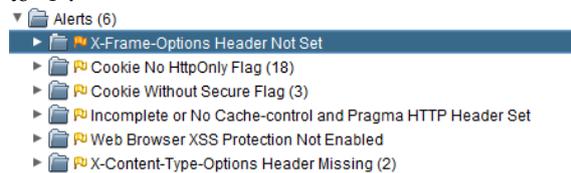

Figure 6b: ZAP Scan Result for Main Server (AWS) of Secure CDP.

**Main Server of Non-Secure CDP**

OWASP scan identified three vulnerabilities in the Main Server of non-secure CDP as shown in Fig – 7a. Identified vulnerabilities are related to password auto-completion, which poses a serious threat to the security of CDP.

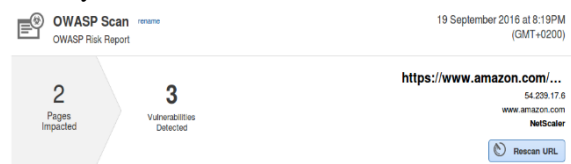

Figure 7a: OWASP Scan Result for Main Server (AWS) of Non-secure CDP.

ZAP scan found around 42 vulnerabilities of eight types in the Main Server of non-secure CDP as shown in Fig – 7b. Most of the vulnerabilities identified are of the same kind as found for Main Server of secure CDP, however, the number of vulnerabilities increased for non-secure CDP. Additionally, as shown by OWASP scan as well, Main Server of non-secure CDP has password auto-completion vulnerability that is a serious issue in relevance to the security of CDP.

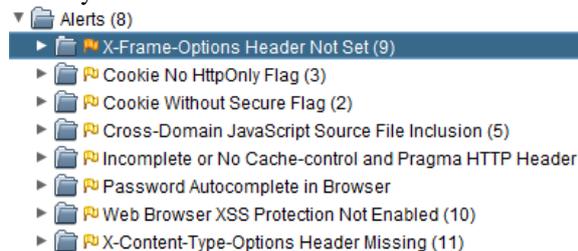

Figure 7b: ZAP Scan Result for Main Server (AWS) of Non-secure CDP.

### 4.2.3 CI Server (Jenkins)

Similar to Main Server, scanning tests are applied on CI servers of both CDPs.

**CI Server of Secure CDP**

OWASP scan did not find any vulnerability in CI server of secure CDP as shown in Fig – 8a.

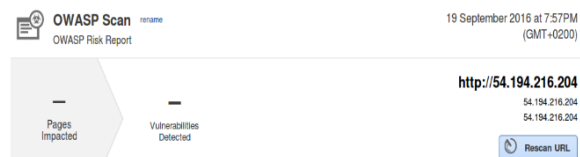

Figure 8a: OWASP Scan Result for CI Server (Jenkins) of Secure CDP.

As shown in Fig – 8b, ZAP scan found around 113 vulnerabilities of five types. The majority of the vulnerabilities are the same as found for the main server and it has already been discussed how these issues can be addressed. A single serious vulnerability is found which relates to password auto-completion.

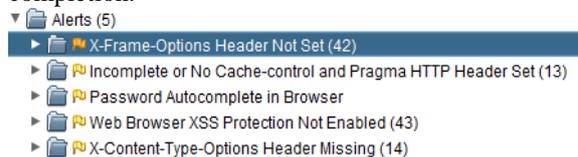

Figure 8b: ZAP Scan Result for CI Server (Jenkins) of Secure CDP.

**CI Server of Non-Secure CDP**

In CI server of non-secure CDP, OWASP scan could not find any vulnerability as shown in Fig – 9a.

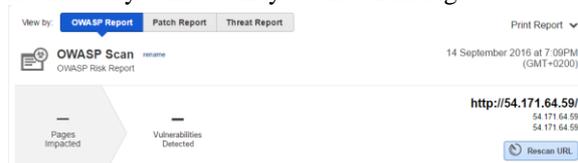

Figure 9a: OWASP Scan Result for CI Server (Jenkins) of Non-secure CDP.

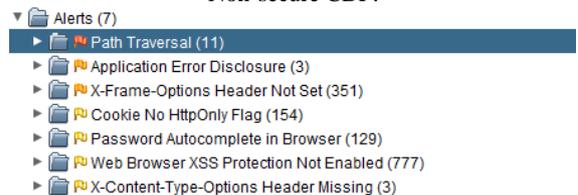

Figure 9b: ZAP Scan Result for CI Server (Jenkins) of Non-secure CDP.

Unlike OWASP scan, ZAP scan found around 1428 vulnerabilities of seven types in CI server of non-secure CDP that is quite a huge number as compared to 113 found for secure CDP (see Fig – 9b). In addition to the vulnerabilities found in CI server of secure CDP, zap found several other serious vulnerabilities in CI server of non-secure CDP.

Table 3: Comparison of vulnerabilities found in components of secure and non-secure CDP.

| | Scan Test | OWASP | | ZAP | | Total | |
|---|---|---|---|---|---|---|---|
| | | Secure CDP | Non-secure CDP | Secure CDP | Non-secure CDP | Secure | Non-secure |
| **Component** | **Main Server** | 3 | 3 | 26 | 42 | 29 | 45 |
| | **CI Server** | 0 | 0 | 113 | 1428 | 113 | 1428 |
| | **GitHub** | 105 | | 10781 | | 10886 | |

These newly identified vulnerabilities are related to path traversal and application error disclosure. The path traversal vulnerability is serious because it allows an attacker to trick the web server and get unauthorized access to sensitive files. Application error disclosure may disclose sensitive information, which can be used to initiate further malicious attacks. Apart from these serious issues, contrary to a single vulnerability of password auto-completion in CI server of secure CDP, ZAP scan found around 129 such vulnerabilities in CI server of non-secure CDP.

## 5 DISCUSSION

As demonstrated, the proposed security tactics are implemented in secure CDP and evaluated, both qualitatively and quantitatively, to find about its effects. The qualitative analysis genuinely specifies that secure CDP is more secure than non-secure CDP because the access to the repository, main server, and CI server is protected through enhanced authentication and authorization techniques. The quantitative findings show that there are vulnerabilities in both the secure CDP and non-secure CDP. Since password auto-completion option exists in web browsers, therefore, the password can be retrieved for both GitHub and Jenkins, which is a serious security issue. However, non-secure CDP contains serious security risks related to accessing cookies through JavaScript, updating Open SSH and showing local IP publicly at GitHub. The findings of the two security scans are summarized in Table – 3. OWASP scan does not show any difference in the number of vulnerabilities but the nature of vulnerabilities found for secure and non-secure CDP is different. Vulnerabilities found by OWASP scan both for Main Server and CI Server are of serious nature and pose a direct threat to the security of CDP while those found for secure CDP are not so serious and are easily fixable. The results shown by ZAP test approves the effectiveness of our devised tactics both qualitatively and quantitatively. First, the number of vulnerabilities found in non-secure CDP is greater than the secure CDP has. Secondly, after investigation, we found that vulnerabilities identified in non-secure CDP are more severe and pose a serious threat to the security of CDP. From the overall results of the two security scans, it can be established that secure CDP is far less vulnerable to malicious attacks as compared to non-secure CDP and so our proposed security tactics sufficiently improve the security of our CDP.

The question can be raised whether these five security tactics affect each other (particularly in a negative way). An analysis of these tactics in relation to each other would give us a clear picture. The repository (GitHub) is isolated from the rest of the setup, so the control over commit and access rights do not have any consequences in relation to other four security tactics. The connection to the main server (AWS) through private-key over SSH does not have any negative effects on other security tactics rather it empowers the security of other components. Similarly, roles on the main server do not affect any other security tactic, though, it interferes with private-key over SSH but these two operate in different realms. The last two tactics are solely related to Jenkins and they do not have any negative consequences in relation to the effectiveness of other security tactics. From this analysis, it can be concluded that devised security tactics can work together and do not affect each other in any negative way. Here, it is important to mention that our evaluation techniques have certain limitations. Assurance case is merely a framework for structuring argumentation, which is supported by claims and quantitative evidence. A deficiency in this technique is that it requires an iterative and opponent-based process to develop an adequate analysis. The results get fully credible only when they can convince our audiences that software is equipped with a reasonable level of security. From the security findings and general information about the scanning tools, it can be deducted that these tools do not cover security issues relevant to OS and low-level Java and it is highly recommendable to identify and address such issues in order to properly assess the security of CDP. It is also worth mentioning that for leveraging full benefits of the devised security tactics, all other essential security measures should be taken into account. For example, firewall setting needs to be correctly setup to help CDP properly utilize incorporated security tactics.

# 6 CONCLUSION

Keeping in view the vast amount of security threats faced by CDP, it is critical to analyse the CDP's security for identifying gaps and devising security strategies to help secure CDP. In this paper, five security tactics are devised to enhance the security of three major components (repository, main server and CI server) of the CDP, which are: (1) controlled access and commit rights for repository; (2) controlled access to AWS instance using private-key over SSH; (3) use of roles on the main server via leveraging AWS IAM; (4) use of VM plug-in for ensuring initial clear state of Jenkins; (5) use of roles on CI server to control access to Jenkins. After devising these security tactics, two CDPs are implemented, secure CDP that incorporates proposed security tactics and non-secure CDP that does not incorporate three of the proposed security tactics. The security of both CDPs is evaluated through qualitative and quantitative methods. The qualitative analysis shows that secure CDP implemented with security tactics is more secure than non-secure CDP. The quantitative analysis also shows a significant improvement in the security level of secure CDP as evident from the number and nature of vulnerabilities found in both CDPs through two different scanning tests.

    The results obtained through quantitative analysis showed some deviation from expected results, which is due to the fact that these penetration tools are specialized for assessing the security of web application. In next step, we plan to develop a framework for assessing the security of the CDPs. We also plan to incorporate our proposed security tactics in a real CDP project and assess their effects on the security aspect of the CDP. In future research, these five security tactics will be transformed into five security patterns by formally describing them according to the standards set by Gang of Four (GoF)[7] team.

---

[7] http://www.blackwasp.co.uk/gofpatterns.aspx